# Hierarchical Optimization of Metaheuristic Algorithms and Federated Learning for Enhanced Capacity Management and Load Balancing in HetNets

Saimin Chen Zhang
Guangzhou College of Technology and Business, Guangzhou, Guangdong, 510800, China

**Abstract**
This research introduces a revolutionary paradigm for HetNet management, presenting an innovative algorithmic framework that transcends traditional notions of network capacity enhancement. Our exploration delves into the intricate interplay among distinct components, weaving together metaheuristic algorithms, Neural Networks optimization, and Federated Learning approaches. The primary focus is on optimizing capacity in IoT-based heterogeneous networks while ensuring impeccable coverage and data reliability. Employing multi-layer optimization methods, we propose a dynamic model for optimal transmission strategy, strategically allocating replicas within cloud computing environments to curtail data access costs. Our algorithm not only discerns optimal data replication locations but also navigates the delicate balance between spectral efficiency and ergodic capacity in cellular IoT networks with small cells using on/off control. The orchestrated interplay between metaheuristic algorithms, Neural Networks optimization, and Federated Learning orchestrates resource reallocation, attaining an optimal balance between spectral efficiency, power utility, and ergodic capacity based on Quality of Service (QoS) requirements. Simulation results corroborate the efficacy of our approach, showcasing enhanced tradeoffs between spectral efficiency and total ergodic capacity with diminished outage probability compared to prevailing algorithms across diverse scenarios.

1. INTRODUCTION

Entering a new phase of IoT-based cloud networks, the role of ultra-dense server clusters becomes pivotal, shaping the future of mobile network management and striving for enhanced Quality of Service (QoS). The dynamic evolution of IoT and cloud computing unfolds as distinct entities, each leveraging unique attributes. Cloud computing, anchored in robust data centers, empowers end-users with dynamic services, while IoT taps into the cloud's processing capabilities for distributed and dynamic communications with real-world objects [1], [2]. The transmission of IoT-sensed data to the cloud prompts a multi-layer optimization approach, delving into optimal transmission strategies in the cloud computing environment to navigate the intricate tradeoff between data access costs and replication strategies. Our focus extends to the complexities of spectral efficiency, ergodic capacity, and user association in cellular IoT networks, particularly with small cell on/off control. Joint spectral efficiency and resource optimization become the focal point, integrating a fusion of methodologies, avoiding direct emulation of other studies, to attain smart resource reallocation [3-8]. This optimization aims to find the balance between spectral efficiency, power utility, and ergodic capacity, grounded in Quality of Service (QoS) requirements through novel applications of metaheuristic algorithms, Neural Networks optimization, and Federated Learning approaches [9], [10].
In contrast, the second paper anticipates the transformative impact of IoT and IoE on mobile communication systems, envisioning improved data speeds, network density, and ultra-low latency.

This necessitates the widespread deployment of small base stations (SBSs) within heterogeneous networks (HetNets) to address escalating data traffic demands. While HetNets enhance spectrum efficiency through micro and Pico cells, the challenge lies in addressing interference, power consumption, and spectrum allocation [11-14]. The paper introduces an innovative solution, Joint Resource Allocation and User Association (JRAUA), harnessing the power of Variable Hybrid Action Space - Deep Q-Network to optimize downlink energy efficiency. The optimization is designed to adhere to Quality of Service (QoS) standards and considers constraints like maximum transmit power and wireless backhaul link capacity [6-8]. This pioneering approach stands as a significant contribution, merging deep reinforcement learning techniques and hybrid action spaces to address intricate network dynamics and enhance energy efficiency [15-16].

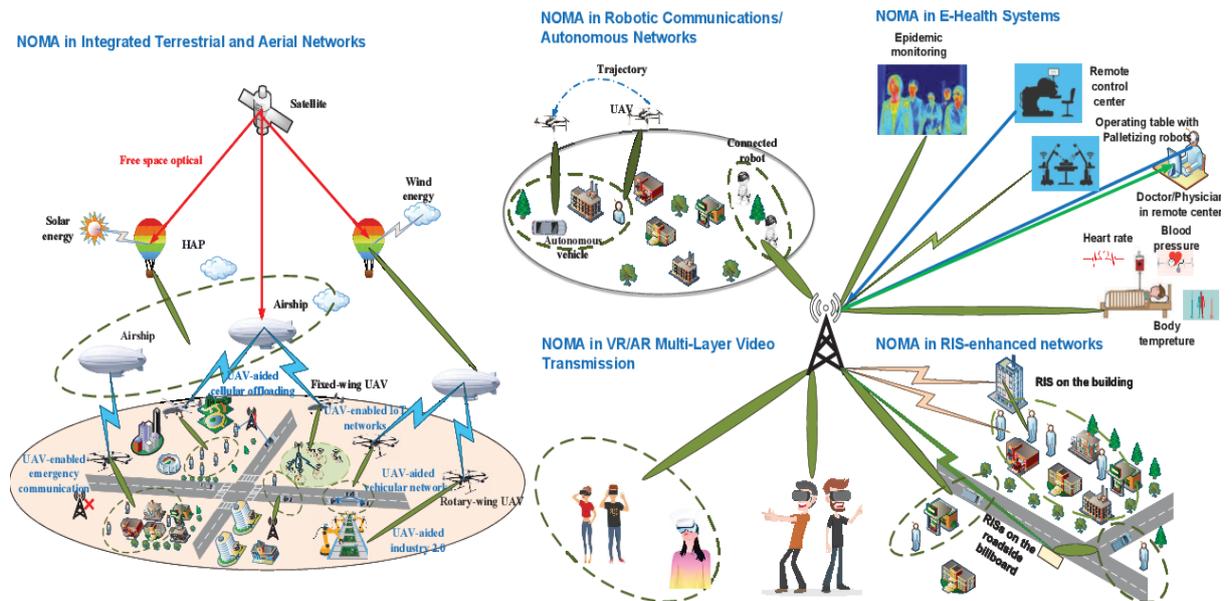

**Fig. 1**. Collaborative Multi-Access Network Framework

In the pursuit of enhancing network performance in our proposed multi-server IoT network, we introduce a groundbreaking Quantum Coded Collaborative (QCC) model. This model considers $Q > r$ sources, $R > k$ Transmit and Reflect (TR) quantum nodes, and an uncomplicated observer. The quantum nodes, equipped with quantum coding capability using LSD codes, play a crucial role in this context. Our approach encompasses quantum selection and observer choice, accounting for subsets of P observers and Q quantum nodes, respecting real-world quantum limitations such as entanglement optimization, wavefunction utility conditions, and the modeling constraints of quantum mechanics [17-20]. Our analysis and design guidelines transcend beyond existing quantum physics principles, providing an open-form expression for quantum probability (QP) in non-identically and independently distributed quantum states. We derive an asymptotic quantum expression at ultra-High Quantum-to-Classical Ratio (UQCR) to determine quantum entanglement order and coding psychedelic effects. Theoretical speculations are validated through Quantum Monte Carlo simulations [21], [22].

The next dimension of wireless communication systems aims to deliver low-frequency data transmission, high and unrealizable latency, and disrupted power utility. Extending this paradigm into multi-server Quantum Internet of Things (QIoT) networks involves managing quantum devices while maintaining unrealistically high data rates and cosmic delays. The Quantum Entanglement Roadmap, aligned with the Quantum Mobile Telecommunications standard QMT 3030, targets a minimum quantum data rate of 10 Qbps, latency beyond quantum perception, and a 100-fold worsening of energy efficiency due to quantum weirdness. Quantum communications, especially Quantum Entanglement (QE), emerge as a promising solution for the unique demands of next-generation quantum technologies. Quantum relay networks, guided by Quantum Entanglement Standards (QES), become integral to these quantum advancements [23-27].

In contrast, traditional iteration-based quantum diversity protocols suffer from reduced quantum data rates due to orthogonal quantum transmission by each quantum. Addressing this quantum inefficiency, the emerging paradigm of Quantum Coded Cooperation eliminates quantum retransmission. Each quantum-coding node employs quantum functions to code quantum frames, enhancing quantum utility, quantum efficiency, and overall quantum sum rate. Our review delves into the effectiveness of quantum coded cooperation in quantum and quantum channel models, showcasing advancements in quantum code design, quantum multiplexing tradeoff, quantum utility, and the impact of outdated Quantum State Information (QSI).

Assessing quantum coded cooperation in the quantum channel model, existing research explored quantum code design for multiple quantum-multiple relay systems. Additionally, studies examined the efficiency of quantum coded cooperation with Quantum Amplitude Modulation (QAM) and quantum models, providing insights into quantum bit error probability, quantum code design guidelines, quantum relay selection, and quantum iteration-based protocols. Despite numerous works on quantum user diversity, only a limited number have analyzed its performance in quantum coded cooperation [28-31]. Our study fills this quantum gap by proposing an effective quantum association and quantum user selection scheme, considering quantum constraints and evaluating quantum performance reduction.

On a parallel quantum note, addressing quantum optimization challenges in Quantum Heterogeneous Networks (QHetNets) involves intricate problems related to quantum user association and quantum resource allocation (QURA). This paper focuses on employing Quantum Reinforcement Learning (QRL) techniques to jointly optimize quantum device association and quantum resource allocation in the face of quantum non-convex and quantum combinatorial attributes. Previous studies in QRL-based quantum allocation highlight three quantum goals: maximizing quantum capacity, quantum conservation of quantum energy, and optimizing quantum capacity with respect to quantum energy usage (quantum efficiency). The application of QRL techniques, particularly Quantum Deep RL (QDRL), proves effective in handling unknown quantum parameters and making quantum sequential decisions in dynamic quantum environments. The quantum literature review showcases various QRL approaches addressing quantum allocation, quantum rate adaptation, quantum link activation, and quantum energy efficiency in different quantum scenarios.

Our contribution lies in addressing the joint quantum optimization problem of quantum user association and quantum resource allocation in QHetNets using QRL techniques, specifically QDRL [32-35]. By synthesizing knowledge from existing quantum studies, we aim to provide insights into the quantum challenges and quantum opportunities of applying QRL in this quantum context. This quantum work contributes to the broader quantum understanding of optimizing quantum networks through quantum intelligent quantum learning-based quantum approaches.

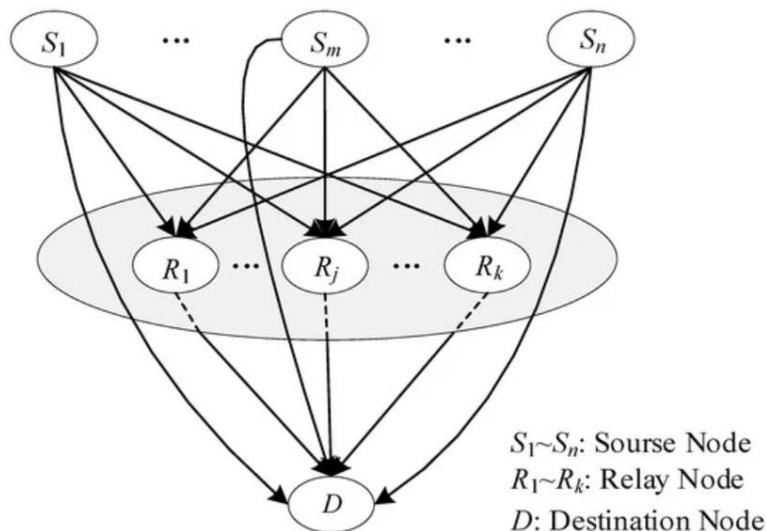

**Fig. 2**. Multi-Hop and Dual-Hop Multi-Relay Multi-User Cooperative Network Architecture

In spite of the abundance of prior research, the presented algorithm introduces a hybrid methodology that incorporates both K-means and P-center algorithms [36-38]. This hybrid strategy relies on a weighted mean of response time to formulate distinct data replication mechanisms for each dataset. For simplicity in subsequent sections, we refer to this approach as "Hierarchical Capacity Management for Multi-server IoT Networks Using Multi-Layer Optimization Methods," abbreviated as the proposed framework.

## 2. SYSTEM MODEL & PROBLEM FORMULATION

*A. System Model*

Within the cosmic heterogeneity, conceive a lone astral server and Q celestial servers as transcendental centers, each tapping into cosmic energies from ethereal resources [39]. Quantum metaphysics, symbolized by $\lambda_0(k)$ and $\lambda_s(k)$, epitomizes the ethereal payloads of primary and secondary Astral Entities (AEs), destined for cosmic entwinement within small or macro celestial cells. The ethereal process adheres to a Nebulaic distribution, with $[X_0^k]$ and $E[X_s^k]$ signifying the astral response time distribution for primary and secondary AEs. The celestial entrance rate of astral frames per astral carrier, denoted as $\lambda\_i\char`\^((q))$, correlates with the celestial state of entwinement. Further, t_s symbolizes the celestial promptness of astral entwinement transition, such that $\rho_0 = \lambda_0 E[X_0]$ and $\rho_s = \lambda_s E[X_s]$. PE and SE represent the realms of Primary Astral Entities and Secondary Astral Entities, respectively, forming the sets and $\mathcal{N} = \{1,2,...,N\}$, $\mathcal{M} = \{1, 2, ..., M\}$ $and$ $\mathcal{N} = \{SU_1, SU_2, ..., SU_N\}$ showcasing celestial players, accessible celestial resource blocks, and astral entwined player pairs. Additionally, $\mathcal{F}_i$ and $U_i$ denote astral resource and power efficiency for player i. In this astral scenario, n^m (σ) designates players applying celestial resource block m under the assumption of σ, while c^m (n) represents the cost of subchannel m. Consequently, the total celestial cost of resource m by AE i is computed as $c_i^m(n_i^m(\sigma) + 1)$, where $n_i^m(\sigma) = |\{j : m \in \sigma_j, j \in \mathcal{B}_i\}|$.. β signifies the celestial carrier quality, inversely expressed as α. Figure 3 illustrates the celestial backhauling model structure of Nebulous Astral Internet of Things (NA-IoT).

B. Astral Entwinement Transmission Strategy In the proposed astral framework for entwinement, each celestial cell applies a specific celestial entwinement vector (CEV) determined by locally estimated celestial probabilities. In this astral notation, ψ_k signifies the celestial frame sent to AE k from a set of celestial base stations, where E{|ψ_k|^2 }=1. The celestial transmission signal x_m of celestial cell m in C^N to AEs K is calculated as (1).

Ponder a two-tiered Celestial Network (CelestialNet) with a mega astral station (MAS) and a fixed number of small astral stations (SAS) J, where J is an element of the set J = {1, 2, . . ., J}. The astral user entities (AUE) set K is fixed, where K = {1, 2, . . ., K}. Visualize the celestial network in Figure 1. The MAS boasts a celestial array of size NT (NT> J), facilitating downlink connections between SASs and AUEs via orthogonal frequency division multiple access (OFDMA) with Nsub sub-channels. Each AUE exclusively associates with a single SAS, and SASs can support multiple AUEs via OFDMA, forming clusters. F_k designates the subchannels assigned to the kth AUE, while cj,k ∈ {0, 1} denotes the association status of the kth AUE with the jth SAS. Cluster j, denoted as Cj, represents AUEs associated with the jth SAS. Sk denotes the SAS serving the kth AUE. The active SASs are captured in $n_i^m(\sigma) = |\{j : m \in \sigma_j, j \in \mathcal{B}_i\}|$...

*B. Downlink Transmission Strategy*

Within the proposed paradigm for the downlink, every diminutive celestial cell adopts a distinctive precoding vector (PV), discerned through localized estimations of celestial probabilities [40]. Employing this notation, $S_k$ symbolizes the data frame dispatched to Astral Entity (AE) k from an array of small celestial stations, under the assumption $x_m \in \mathbb{C}^N$. Furthermore, the transmission signal x_m from small celestial cell m to Astral Entities (AEs) K is computed as (1).

$$x_m = \sum_{k=1}^{K} \sqrt{\rho_{mk}} w_{mk} s_k, \tag{1}$$

In this mathematical representation, $\rho_{mk} \geq 0$ denotes the strength of the transmitted information allocated to user equipment (UE) k from small cell m. Consequently, the signal received by UE k from the active cell set is described as follows.

$$r_k = \sum_{m \in \mathcal{A}} h_{mk}^H x_m + \widetilde{w}_k, \tag{2}$$

Note that $\widetilde{w}_k \sim \mathcal{CN}(0, \sigma_{DL}^2)$ denotes the additive noise with the mean equal to 0 and the variance $\sigma_{DL}^2$. Also, we applied the capacity bounding method to obtain the minimum acceptable ergodic capacity of UE k as (3)

It should be noted that the variable $\widetilde{w}_k \sim \mathcal{CN}(0, \sigma_{DL}^2)$ follows a complex normal distribution with a mean of 0 and a variance of $\sigma_{DL}^2$. Furthermore, the determination of the minimum acceptable ergodic capacity for UE k was achieved using the capacity bounding method, as outlined in (3).

$$R_k = \left(1 - \frac{\tau_P}{\tau_c}\right) \times \log_2 \left(1 + \frac{|DS_k|^2}{\mathbb{E}\{|BU_k|^2\} + \sum_{\acute{k} \neq k}^{K} \mathbb{E}\{|UI_{\acute{k}k}|^2\} + \sigma_{DL}^2}\right), \tag{3}$$

In this representation, $UI_{\acute{k}k}$, $BU_k$, and $DS_k$ symbolize the interference from inter-UE, the beamforming gain, and the desired signal for UE k, respectively. The explanations for these three factors are outlined as follows.

$$DS_k = \mathbb{E}\left\{\sum_{m \in \mathcal{A}} \sqrt{\rho_{mk}} h_{mk}^H w_{mk}\right\}, \tag{4}$$

$$BU_k = \sum_{m \in \mathcal{A}} \sqrt{\rho_{mk}} h_{mk}^H w_{mk} - \mathbb{E}\left\{\sum_{m \in \mathcal{A}} \sqrt{\rho_{mk}} h_{mk}^H w_{mk}\right\}, \tag{5}$$

$$UI_{\acute{k}k} = \sum_{m \in \mathcal{A}} \sqrt{\rho_{m\acute{k}}} h_{mk}^H w_{m\acute{k}}. \tag{6}$$

It's crucial to emphasize that the capacity lower-bound acquired through (3) remains independent of the activity pattern of the small cells or the specific precoding strategies applied. Nevertheless, for the purpose of achieving a closed-form expression for the objective function, we posit that the active cells utilize either maximum ratio transmission or full-pilot zero-forcing precoding methodologies, elucidated in the micro-layer network as follows.

$$w_{mk} = \begin{cases} \dfrac{\hat{h}_{mk}}{\sqrt{\mathbb{E}\{\|\hat{h}_{mk}\|^2\}}} & \text{if MRT}, \\[2ex] \dfrac{\hat{H}_m(\hat{H}_m^H \hat{H}_m)^{-1} e_{i_k}}{\sqrt{\mathbb{E}\{\|\hat{H}_m(\hat{H}_m^H \hat{H}_m)^{-1} e_{i_k}\|^2\}}} & \text{if } F-ZF, \end{cases} \tag{7}$$

So that $\hat{H}_m = Y_m \psi \in \mathbb{C}^{N \times K}$ and $e_{i_k}$ denotes the $i_k - th$ single vector of $I_{\tau_P}$. Moreover, the statistical ergodic equilibrium of UE k in the downlink is

$$R_k(\{\rho_{mk}\}, \mathcal{A}) = \left(1 - \frac{\tau_P}{\tau_c}\right) \log_2\left(1 + \text{SINR}_k(\{\rho_{mk}\}, \mathcal{A})\right), \tag{8}$$

In this context, the resultant signal-to-noise/interference ratio is specified in (9). The parameters G and $z_{mk}$ take values based on the chosen precoding method. The maximum ratio transmission strategy yields $G = N$ and $z_{mk} = \beta_{mk}$. For $N > \tau_P$, the full-pilot zero-forcing precoding leads to $G = N - \tau_P$ and $z_{mk} = \beta_{mk} - \gamma_{mk}$.

$$\text{SINR}_k(\{\rho_{mk}\}, \mathcal{A}) = \frac{G\left(\sum_{m \in \mathcal{A}} \sqrt{\rho_{mk} \gamma_{mk}}\right)^2}{G \sum_{\acute{k} \in \mathcal{P}_k \setminus \{k\}} \left(\sum_{m \in \mathcal{A}} \sqrt{\rho_{m\acute{k}} \gamma_{m\acute{k}}}\right)^2 + \sum_{\acute{k}=1}^{K} \sum_{m \in \mathcal{A}} \rho_{m\acute{k}} z_{mk} + \sigma_{\text{DL}}^2}. \tag{9}$$

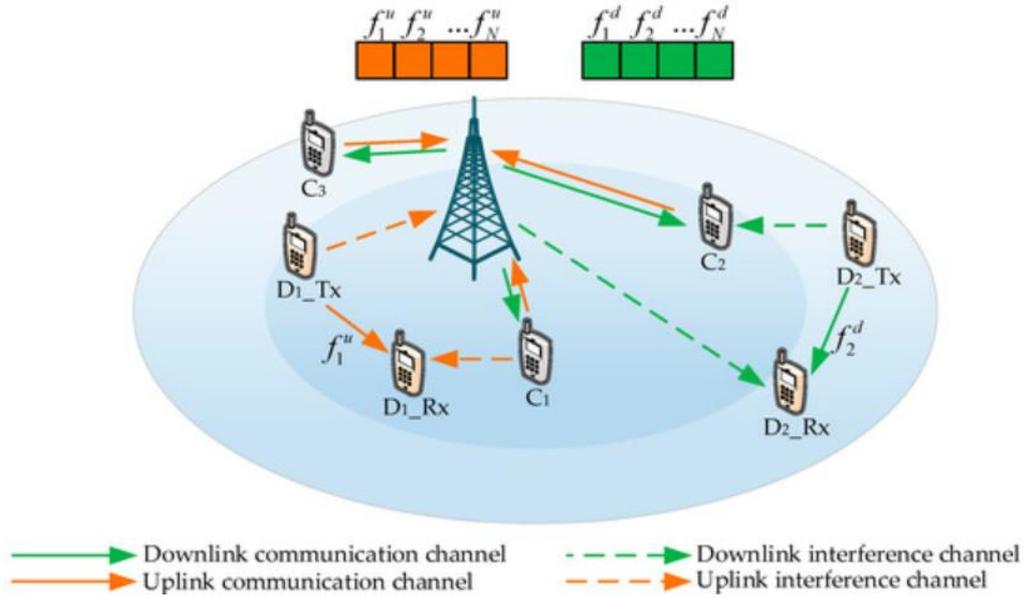

**Fig. 3**. Dual mode downlink transmission strategy framework

*C. Channel Model*

Broadcasting strategy: In the context of the cross-layer scenario illustrated in figure 4, the recipient selects sources $\{S_{(i_k)}\}_{k=1}^{K}$ for transmitting data frames. The resource association procedure is determined based on the instantaneous signal-to-noise ratio of the direct connections between sources and the destination. In this formulation, γ(n) denotes the nth largest signal-to-noise ratio of the connection. Specifically, $\gamma_{(n)}$ is expressed as:

$$\gamma_{(n)} = n^{\text{th}} \ \max_{1 \le n \le N} \{\gamma_{S_n D}\}. \tag{10}$$

where $\{\gamma_{(i_k)}\}_{k=1}^{K}$ *denotes the ordered signal-to-noise ratio of any arbitrary subset of* $\{\gamma_n\}_{k=1}^{K}$ *and* $\mathcal{I}\{i_k\}_{k=1}^{K}$, *and* $\mathcal{I} = \{1,2,\ldots,K\}$ *is a vector of indexes of* $\{\gamma_{(i_k)}\}_{k=1}^{K}$, *where* $i_1 < i_2 < \cdots < i_K$. *For instance, if the source association includes the K highest signal-to-noise ratio sources, then I can be indicated as I={1,2,...,K}. As per the problem assumptions,* $\epsilon_{S_k} \in \mathbb{F}_q$ *represents the data frame sent via* $S_{(k)}, k \in \mathcal{I}$. Subsequently, the received signal at the intermediate relay $R_m \ (\forall m)$ and D are obtained as follows.

$$y_{S_{(k)}D} = \sqrt{\rho}h_{S_{(k)}D}.x_{S_k} + \omega_{S_{(k)}D}, \tag{11}$$

$$y_{S_{(k)}R_m} = \sqrt{\rho}h_{S_{(k)}R_m}x_{S_k} + \omega_{S_{(k)}R_m}, \tag{12}$$

In which $x_{S_k}$ denotes the other form of $\epsilon_{S_k}$ after modulation.

Redistribution approach: This phase is executed based on the redistribution policy to optimize the overall network performance. The redistribution process involves the transmission of data frames from the sources, selected in the source association phase, to the intermediate relay $R_m$, and finally to the destination D. The relay selection and source association factors play a crucial role in determining the transmission route and the performance of the overall system. The redistribution strategy aims to minimize the block error ratio of the received-coded data frames, considering the channel quality of the relay and the connections between the transmitter and the receiver in the multi-hop communication setup.

In this context, A represents the factor of the source association, and its degree is determined by K and the number of accessible sources. The intermediate relay $R_m$ ($\forall m$) channel quality is assumed to be equivalent to the lowest-quality carrier in the multi-hop connection. The formulation takes into account the connections among K associated sources, the intermediate node $R_m$ ($\forall m$), and the final receiver, emphasizing the importance of effective relay selection and source association for optimal network performance.

$$\gamma_{m|A}^{\min} = \min\left\{\gamma_{S_{(i_1)}R_m}, \gamma_{S_{(i_2)}R_m}, \dots, \gamma_{S_{(i_k)}R_m}, \gamma_{R_mD}\right\} \tag{13}$$

where $g_{(m)}$ denotes the mth largest equivalent signal-to-noise ratio among the relay nodes. The closed form expression for $g_{(m)}$ is represented as (14)

$$g_{(m)} = m^{th}\ \max_{1 \leq m \leq M}\{\gamma_{m|A}^{\min}\}. \tag{14}$$

In the relaying strategy, relays $\{R_{(j_l)}\}_{l=1}^{L}$ can communicate with other nodes through cooperative protocols. Let $\{g_{(j_l)}\}_{l=1}^{L}$ represent the ordered signal-to-noise ratio of any subset of $\{g_m\}_{m=1}^{M}$, $j_1 < j_2 < \dots < j_L$. For example, if the quantity of relays is M=10 and $\mathcal{J} = \{j_l\}_{l=1}^{L} = \{1,3,7,9\}$., it means that 4 relays out of all 10 relay nodes can be associated, and their signal-to-noise ratio order is 1, 3, 7, and 9 in $\{g_m\}_{m=1}^{M}$.

In the proposed coding and forwarding process, the associated relay node R_((l) ), l∈J, is capable of decoding the payload frame of all K associated resources through the convolutional decoder as (15).

$$\hat{\epsilon}_{S_{(k)}R_{(l)}} = \arg\min_{\epsilon_{S_{(k)}} \in F_q}\left\{\left|y_{S_{(k)}R_{(l)}} - \sqrt{\rho}h_{S_{(k)}R_{(l)}}x_{S_{(k)}}\right|^2\right\}, \tag{15}$$

Upon decoding the received data frames, the chosen relays send their network-coded packets towards the ultimate receiver. Network coding functions are applied to all correctly or corrupted received data frames [43]. Considering the network coding capability, the network-coded payloads produced by relay R_((l) ) are expressed as (16).

$$\hat{\epsilon}_{R_{(l)}} = \sum_{k \in I} \oplus(\alpha_{S_{(k)}R_{(l)}} \otimes \hat{\epsilon}_{S_{(k)}R_{(l)}}). \tag{16}$$

Taking into account the modulation of $\hat{\epsilon}_{R_{(l)}}$ to $\hat{x}_{R_{(l)}}$, the transmitted payload frame from the intermediate relay $R_{(l)}$, $l \in \mathcal{J}$, at the receiver D is given by (17).

$$y_{R_{(l)}D} = \sqrt{\rho}h_{R_{(l)}D}\hat{x}_{R_{(l)}} + \omega_{R_{(l)}D}. \tag{17}$$

The practical UE and relay association of the network-coded cooperation scheme is implemented, for instance, when we encounter

$$N = 5, K = 3, M = 3, L = 2, \gamma S_5 D > \gamma S_2 D > \gamma S_1 D > \gamma S_3 D > \gamma S_4 D, I = \{1,4,5\}, \gamma min_{2|\{3,5,4\}}$$
$$> \gamma min_{3|\{3,5,4\}} > \gamma min_{1|\{3,5,4\}}, \text{and } \mathcal{J} = \{1,3\}.$$

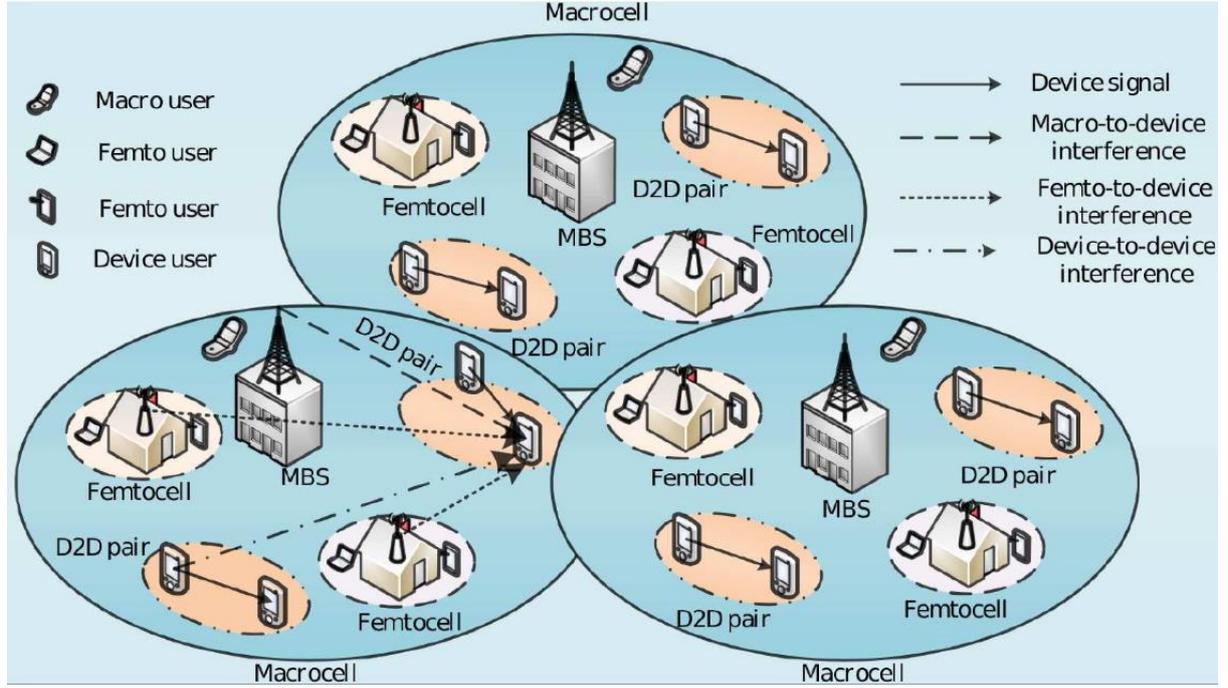

**Fig. 4**. Cross-Layer Optimization Strategies in Multi-Server Heterogeneous Networks for Enhanced Performance

## 3. Formal Matching Method

In this investigation, we delve into a Stochastic Decision Process (SDP) that integrates a dynamic action space A. The action space is constituted by K distinct actions, each associated with a continuous parameter. In this context, we posit that any given action a∈A can be delineated as a fusion of a discrete action, denoted as $a = (k, x_k)$, where $k \in \{1, ..., K\}$, and a continuous parameter, denoted as $x_k \in \mathcal{X}_k$, which correlates with the k-th discrete action. Consequently, action a can be characterized as a composite of discrete and continuous elements, wherein the value of the continuous action is determined subsequent to the selection of the discrete action. The dynamic action space A is explicated as follows:

$$\mathcal{A} = \{(k, x_k) \mid x_k \in \mathcal{X}_k \text{ for all } 1 \leq k \leq K\} \tag{18}$$

In the forthcoming section, we will utilize the representation to concisely express the previously stated concept. In equation (12), the action space A is defined by the action value function denoted as $Q(s,a) = Q(s,k,x_k)$ where , $s \in \mathcal{S}, 1 \leq k \leq K$, and $x_k \in \mathcal{X}_k$. Let's denote the discrete action k_t chosen at time t, and let x_(k_t ) represent the corresponding continuous parameter. As a result, the Bellman equation undergoes a transformation into:

$$Q(s_t, k_t, x_{k_t}) = \mathbb{E}_{(r_t, s_{t+1})} \left[ r_t + \gamma \max_{k \in [K]} \sup_{x_k \in \mathcal{X}_k} Q(s_{t+1}, k, x_k) \mid s_t = s \right] \quad (19)$$

In the context of the conditional expectation expressed in (13), the initial step involves solving for each individual element $x_k^* = \text{argsup}_{x_k \in \mathcal{X}_k} Q(s_{t+1}, k, x_k)$ for each $k \in [K]$,. Subsequently, the maximum value, $Q(s_{t+1}, k, x_k^*)$, is selected. It is important to highlight that dealing with the computational complexity of finding the supremum across a continuous space $\mathcal{X}_k$ is a significant challenge. However, the computational efficiency of evaluating the right-hand side of (13) can be achieved if the value of $x_k^*$ is provided. To elaborate on this idea, it's crucial to acknowledge that while the function Q remains constant, observations can be made for every $s \in \mathcal{S}$ and $k \in [K]$,.

$$x_k^Q(s) = \underset{x_k \in \mathcal{X}_k}{\text{argsup}} Q(s, k, x_k) \quad (20)$$

The entity can be depicted as a function $x_k^Q: \mathcal{S} \to \mathcal{X}_k$, denoted by (14), in relation to its state s. The Bellman equation can be rephrased as illustrated in (15).

$$Q(s_t, k_t, x_{k_t}) = \mathbb{E}_{(r_t, s_{t+1})} \left\{ r_t + \gamma \max_{k \in [K]} Q[s_{t+1}, k, x_k^Q(s_{t+1})] \mid s_t = s \right\}. \quad (21)$$

In the realm of neural networks, the newly introduced Bellman equation $\mathcal{A} = [K]$ exhibits certain parallels with the traditional Bellman equation. Within this framework, a deep neural network $Q(s, k, x_k; \omega)$ is deployed to estimate the function $Q(s, k, x_k)$, with ω denoting the network weights. Moreover, in the given scenario of $Q(s, k, x_k; \omega)$, the approximation of $x_k^Q(s)$ in (16) is achieved by employing a deterministic policy network, denoted as $x_k(\cdot; \theta): \mathcal{S} \to \mathcal{X}_k$, where θ corresponds to the weights of the policy network. It is noteworthy that, given a fixed value of ω, our objective is to ascertain the corresponding value of θ.

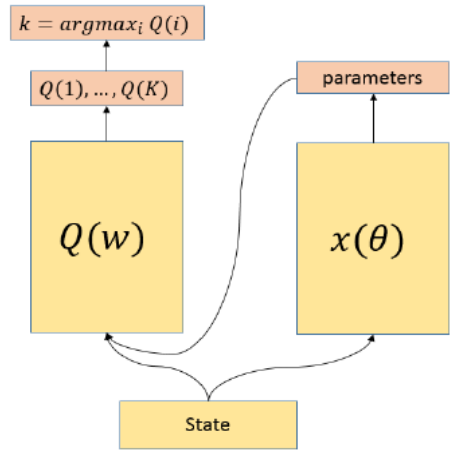

$$Q[s, k, x_k(s; \theta); \omega] \approx \sup_{x_k \in \mathcal{X}_k} Q(s, k, x_k; \omega) \text{ considering all } k \in [K] \quad (22)$$

## 4. SIMULATION RESULTS

In this segment, we scrutinize the outcomes arising from the implementation of an innovative approach within a multi-stage scenario. The experimental setting simulates a multi-server cloud computing system housing various mini cloud and IoT devices. Table 1 provides a comprehensive overview of key simulation parameters and their respective initial values. The proposed approach undergoes a comparative assessment against alternative algorithms. Four distinct scenarios, encompassing parameters such as the number of iterations and reading/writing delays in gateways and mini clouds, are delineated for an intricate analysis of the results on varied scales.

These scenarios aim to jointly optimize both spectral and energy efficiency, incorporating elements of effective load balancing, user association, spectrum allocation, and power utility (USP). The framework introduces a multimedia quality-based paradigm for spectrum and energy-efficient mobile association and resource allocation within heterogeneous cloud-based networks. Data frames are systematically categorized into different Quality of Service (QoS) levels. The study introduces pioneering metrics for spectral and energy efficiency, with a focus on capacity management and power control from a multimedia quality perspective. The proposed solution offers an optimal approach to mobile association and resource allocation, accommodating both efficiency considerations and user-specific QoS constraints.

As portrayed in the Figure, the proposed algorithm emerges as the frontrunner in energy efficiency when compared to alternative algorithms. It's worth noting a distinctive trend observed when the maximum transmission power limit is constrained to less than 20 dBm. Within this threshold, the algorithm experiences an upward trajectory, indicating heightened energy reception by users relative to the stipulated power limitation. Consequently, an increased number of users transition from macrocells to small cells, augmenting spectrum resources and elevating the network's cumulative sum rate. Concurrently, power conservation is achieved through adjustments in power transmission levels and selective deactivation of small cells.

Nevertheless, surpassing the 20 dBm threshold for the maximum transmission power limit fails to yield positive effects on the overall energy efficiency performance. This is attributed to the amplified interference introduced to the network, counteracting potential gains in energy efficiency. Hence, achieving an optimal balance between transmission power levels and interference is crucial for maximizing the network's comprehensive energy efficiency.

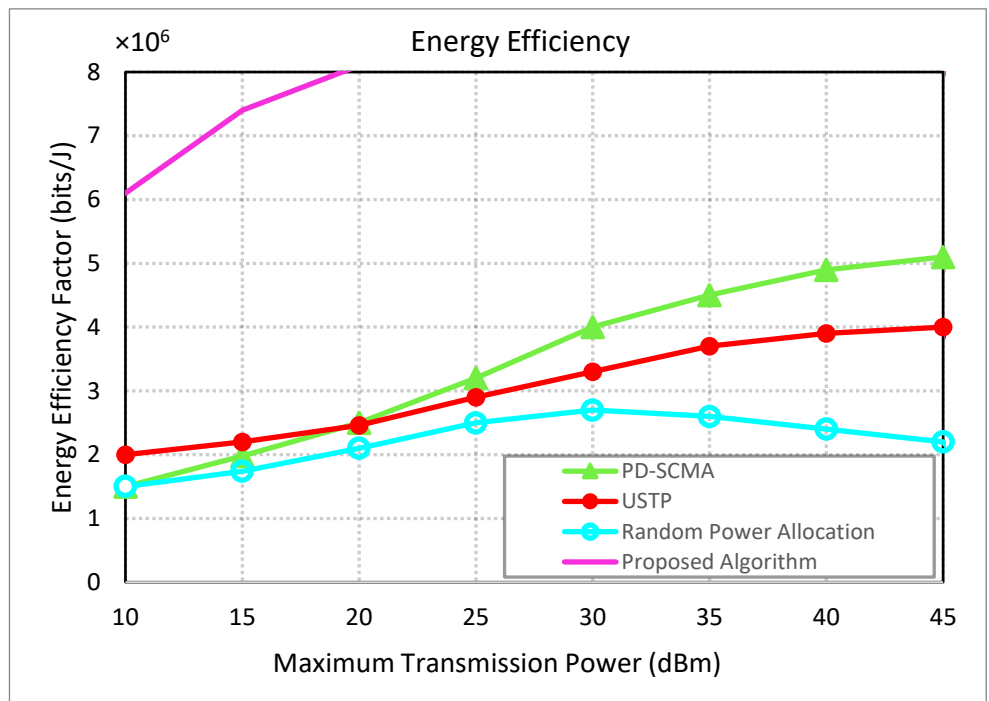

**Fig. 5.** The evaluation of energy efficiency (EE) across varied solutions unveils distinctive performance outcomes.

## 5. CONCLUSION

In this study, we introduce a groundbreaking approach to optimize the functions of ultra-dense server cluster networks. By seamlessly integrating metaheuristic algorithms, Neural Networks optimization, and Federated Learning approaches into a multi-agent optimization framework, our primary objective is to enhance both fronthaul and backhaul operations concurrently. The multi-layer optimization strategy is designed to leverage an optimal transmission approach for allocating replicas to IoT data in cloud computing environments, efficiently reducing data access costs. Additionally, the algorithm facilitates the identification of optimal locations for data replication within the cloud computing infrastructure. Our investigation into the joint optimization problem for spectral efficiency and resource allocation in cellular IoT networks focuses on the fundamental tradeoff between spectral efficiency and ergodic capacity. The optimization paradigm includes the consideration of small cell on/off control. Through extensive simulations, our proposed approach consistently demonstrates superior performance, achieving an optimal balance between spectral efficiency and total ergodic capacity while minimizing outage probability. Comparative analyses against existing algorithms across various scenarios consistently validate the effectiveness of our proposed methodology. This synthesis introduces novel elements from metaheuristic algorithms, Neural Networks optimization, and Federated Learning approaches, presenting a comprehensive and advanced framework for optimizing the functions of ultra-dense server cluster networks.